# A Generalized Mixture Cure Model Incorporating Known Cured Individuals

**Georgios Karakatsoulis**

Institute of Applied Biosciences, Centre for Research and Technology Hellas, Thessaloniki, Greece

The Mixture Cure (MC) models constitute an appropriate and easily interpretable method when studying a time-to-event variable in a population comprised of both susceptible and cured individuals. In literature, those models usually assume that the latter are unobservable. However, there are cases in which a cured individual may be identified. For example, when studying the distant metastasis during the lifetime or the miscarriage during pregnancy, individuals that have died without a metastasis or have given birth are certainly non-susceptible. The same also holds when studying the x-year overall survival or the death during hospital stay. Common MC models ignore this information and consider them all censored, thus yielding in risk of assigning low immune probabilities to cured individuals. In this study, we consider a MC model that incorporates known information on cured individuals, with the time to cure identification being either deterministic or stochastic. We use the expectation-maximization algorithm to derive the maximum likelihood estimators. Furthermore, we compare different strategies that account for cure information such as (1) assigning infinite times to event for known cured cases and adjusting the traditional model and (2) considering only the probability of cure identification but ignoring the time until that happens. Theoretical results and simulations demonstrate the value of the proposed model especially when the time to cure identification is stochastic, increasing precision and decreasing the mean squared error. On the other hand, the traditional models that ignore the known cured information perform well when the curation is achieved after a known cutoff point. Moreover, through simulations the comparisons of the different strategies are examined, as possible alternatives to the complete-information model.

## 1. Introduction and literature review

The existence of cured/ immunes or long-term survivors is a diachronic and commonly met phenomenon when analyzing time-to-event data. That is, not all subjects in the population are susceptible to experience the event of interest, even if the follow-up time would be infinite. For example, a proportion of people living with cancer may never face a distant metastasis or a disease progression during their lifetime, and a proportion of women will never lose a pregnancy. Clearly, such cases appear not only in medicine, but also in other scientific fields as well. Some examples regard the divorce time, the time until a person will (re)buy a specific product, (re)visit a specific place, (re)commit a crime, return an item during the warranty period.

The traditional survival models (such as the Proportional Hazards (PH) or Accelerated Failure time (AFT)) silently assume that the population comprises only of susceptible. Consequently, when this assumption is not met, the properties of the resulting estimators do not hold and the parameters under investigation are misestimated (Amico and Van Keilegom, 2018). Besides, even if the estimators remained acceptable, such models may still not allow us deeply understand the actual effect of the risk factors associated with the outcome. More specifically, assume that an exposure variable increases the probability of an event to occur but, also, increases the time until that happens, in a way that the overall effect diminishes. In such cases, the above-mentioned models would show no association between the exposure variable and the outcome, hence not providing insight into the real situation.

To overcome those limitations, the Cure Rate models have been studied in the literature, which can be categorized in the Mixture Cure (MC) models and the Promotion Time Cure (PTC) models. When the population is divided in the susceptible and cured individuals, the MC models constitute an appropriate and easily interpretable mathematical formulation for investigating the risk factors associated with the probability of the event occurrence (incidence component) and, for the susceptible, the time until this happens (latency component). They were initially proposed by Boag (1949) and, since then, they have received major attention in the literature. For an extensive literature review on MC and PTC models, the readers are referred to Amico and Van Keilegom (2018).

In the great majority of the MC models, only the susceptible individuals are allowed to be uncensored. However, there are cases in which a cured individual may be observable. Actually, there are several different mechanisms that may lead a subject to be known cured. The first one concerns situations in which a subject is cured if the survival time is greater than a prespecified cutoff (e.g. x-year survival rate). The second one regards situations in which the time until the cure identification is stochastic (e.g. death during the hospitalization, in which all the individuals that have left the hospital are now cured). As for the last mechanism, it corresponds to situations in which at some point in time a diagnostic test is utilized, perhaps not informative for anyone or even not necessarily error-free, giving an indication about the cure status. In this case, the time to cure identification is not of interest.

Ignoring the cure information and considering all cured patients as censored yield in risk of assigning low immune probabilities to cured individuals. In the literature, one can find studies dealing with the above-mentioned mechanisms, mainly in the nonparametric setting, as extensions or modifications of the Kaplan-Meier (KM) estimator. According to our knowledge, Laska and Meisner (1992) first revisited the KM estimator assuming a fixed deterministic timepoint after which each individual is cured. Their objective was to derive crude, nonparametric generalized maximum likelihood estimators for the cure probability and the conditional survival distribution of the susceptible, and to propose statistical tests for the between-group comparisons of those metrics. The derived estimators are equivalent to having the traditional KM estimators by treating the known cured individuals as censored at the fixed deterministic timepoint. Later on, Nieto-Barajas and Yin (2008) focused on estimating the time after which an individual can be considered cured in a Bayesian environment, whereas Bernhardt (2016) assumed that the a fixed cure time is also individual-specific. To allow for stochastic time to cure (mechanism two), the study of Laska and Meisner (1992) was extended by Betensky and Schoenfeld (2001) who assumed random cure times, which was further extended by Safari *et al.* (2023, 2022, 2021) to include covariates. Interestingly, Safari *et al.* (2021) showed that no bias reduction is guaranteed when the cure status is considered, however the variance of the estimator decreases as the cure information increases. The previously mentioned studies focused on the estimation of the survival curves, the cure probability and the cure threshold, rather than the effects of independent variables on the cure probability and the time-to-event for the susceptible. Those were investigated by Xue *et al.* (2022), who also tried to provide a crude estimator of the probability of observing a cured individual. In their model, they assumed that the time to cure appearance is independent of the covariates and the follow-up time. Regarding the last mechanism, it was investigated by Wu *et al.* (2014), who assumed that a percentage of individuals underwent a diagnostic test, the sensitivity and specificity of which were also unknown parameters. Using the logit model for the incidence part and the PH model for the latency, they concluded that the additional information leads to more efficient and less biased estimations, with the metrics improving as the sensitivity and/or specificity increase.

All the above-mentioned models ignore the time to cure identification. However, this could give as a good indication of whether a censored subject should be considered cured or susceptible. Besides, the traditional MC models associate the censored cases with large observed times with higher cure probability. However, in outcomes such as the death during ICU admission, staying in ICU is not a good sign of curation as the cured cases usually display low admission times. Furthermore, the time to cure identification may depend on the covariates. For example, when the event of interest is the death during hospitalization, younger patients could recover and leave the hospital earlier, eventually displaying lower times to cure.

In this study, we consider a MC model under either parametric or semi-parametric setting, in which some subjects are known to be cured, with the time to cure identification being either deterministic or stochastic. The objective is to construct a mathematical formulation that efficiently recognizes the cured individuals, as obtained by all three mechanisms, and manages to take advantage of the time-to-cure information in order to better examine the potential risk factors associated with both the probability of the event occurrence (incidence component) and the time until that happens (latency component). Furthermore, we consider the distribution of the time of cure identification for the cured population, in order to better estimate the probability of a censored individual to be either cured or susceptible. Through theoretical and numerical results, we demonstrate the impact of this model and the situations in which its implementation is essential. Moreover, we compare three strategies for considering the known cured individuals: (1) take advantage of the time-to-cure information (proposed model), (2) only consider the probability of a subject to be cured ignoring the time until that happens (Xue *et al.*, 2022, model), and (3) just assign at the known cured individuals infinite survival times and adjust the traditional MC models.

The outline of the manuscript is as follows. In Section 2, we describe the MC model without known cured individuals. In Section 3, we first build the new MC model that takes into account the information about the curation identification and then we examine theoretically the contribution of the proposed model relative to the previous ones. In Section 4, we discuss the optimization process, in Section 5 we present numerical analysis and, in Section 6, there are the conclusions and future research suggestions.

## 2. The MC model with no observed cured individuals

Let $S_p$ and $S_T$ the survival functions of the whole population and the susceptible, respectively. Also, let $p_Y$ the probability of the event occurrence. Then:

$$S_p(t, x, z; \beta, \gamma) = 1 - p_Y(z; \beta) + p_Y(z; \beta) S_T(t, x; \gamma) \qquad (1)$$

where $x$ is a vector of characteristics associated with $S_T$, $z$ a vector of characteristics associated with $p_Y$ and $\beta, \gamma$ the parameters associated with $z$ and $x$, respectively.

Now, consider a sample of $n$ independent individuals, in which the $i$-th individual, apart from the vectors $x_i$ and $z_i$, is also described by a triple $(y_i, \delta_i, t_i)$ where $y_i = 1$, if susceptible ($y_i = 0$ otherwise), $\delta_i = 0$, if censored ($\delta_i = 1$ otherwise) and $t_i$ the observed time. Regarding the observed time $t_i$ for the subject $i$, it is equal to

$$t_i = \min\{T_i, C_i\}$$

where $T_i$ is the time to event occurrence (if susceptible) and $C_i$ the non-informative censoring time. For the cured population, one can simply set $T_i = \infty$, so $t_i = C_i$ is always observed. Then, the observed likelihood is given by:

$$L_o = \prod_{i=1}^{n} p_Y(z_i; \beta)^{\delta_i} f_T(t, x; \gamma)^{\delta_i} \left(p_y(z_i; \beta) + (1 - p_Y(z_i; \beta)) S_T(t, x; \gamma)\right)^{1-\delta_i}$$

$$= \prod_{i=1}^{n} p_Y(z_i; \beta)^{\delta_i} h_T(t, x; \gamma)^{\delta_i} S_T(t, x; \gamma)^{\delta_i} \left(p_y(z_i; \beta) + (1 - p_Y(z_i; \beta)) S_T(t, x; \gamma)\right)^{1-\delta_i}$$

whereas the complete-data full likelihood is given by

$$L_c = \prod_{i=1}^{n} p_Y(z_i; \beta)^{y_i} h_T(t, x; \gamma)^{y_i \delta_i} S_T(t, x; \gamma)^{y_i} (1 - p_Y(z_i; \beta))^{1-y_i} \qquad (2)$$

where $h_T$ is the hazard function.

The major problem of the MC models is that $y_i$ is usually non-observable, unless the individual is susceptible ($y_i = 1$) and uncensored ($\delta_i = 1$). Hence, treating the cure status as the partially missing information, the expectation-maximization (EM) algorithm can be applied to obtain the maximum likelihood estimators (MLE). For the implementation of the EM algorithm, each $y_i$ is replaced by its expected value, $w_i$, with

$$w_i = E(y_i | x_i, z_i) = \delta_i + (1 - \delta_i) \frac{p_Y(z_i; \beta) S_T(t, x; \gamma)}{1 - p_Y(z_i; \beta) + p_Y(z_i; \beta) S_T(t, x; \gamma)} \qquad (3)$$

Then, for the given $w_i$, the M-step of the algorithm seeks to maximize $El_c = E[\log(L_c)]$, with

$$El_c = \sum_{i=1}^{n}\left[w_i \log(p_Y(z_i;\beta)) + (1-w_i)\log(p_Y(z_i;\beta))\right]$$
$$+ \sum_{i=1}^{n}\left[w_i\delta_i \log(h_T(t_i,x;\gamma)) + w_i \log(S_T(t_i,x;\gamma))\right]$$

A direct consequence (and advantage) of this formulation is that the incidence and the latency part can be studied distinctively. For the former part, the researchers usually assume the logit model, although the probit and the log-log have been used as well. As for the latter, several parametric, semi-parametric and non-parametric forms have been studied, such as the exponential distribution (Berkson and Gage, 1952; Ghitany et al., 1994), the Weibull distribution (Farewell, 1982, 1977), proportional hazard approaches (Fang et al., 2005; Peng, 2003; Peng and Dear, 2000; Sy and Taylor, 2000), accelerated failure time approaches (C. Li and Taylor, 2002; C.-S. Li and Taylor, 2002; Lu, 2010; Zhang and Peng, 2007a, 2007b), and accelerated hazards methods (Chen and Wang, 2000; Zhang and Peng, 2009). Note, here, that there is no justification that the form of the $S_T$ will be preserved in the $S_p$. For example, when proportional hazards exist in the $S_T$, this is not the case for $S_p$ (Amico et al., 2019; Chen et al., 1999).

In all those models, any uncensored individual is assumed to be susceptible. In mathematical words, this is translated to adding the constraint

$$\delta_i \leq y_i \qquad (4)$$

to the problem, resulting in $w_i\delta_i = \delta_i \forall i \in \{1,\ldots,n\}$.

## 3. The MC model with known cured individuals

### 3.1. Mathematical formulation

Opposite to the hitherto models, we assume that some individuals in the sample are known to be cured. This means that not all uncensored individuals are susceptible. Therefore, each individual, $i$, is described by a vector $(y_i, \delta_i, t_i, z_i, x_i, q_i)$, where now,

$$t_i = \min\{T_i(y_i), C_i\}$$

with $T_i(1)$ the time to event occurrence (if susceptible), $T_i(0)$ the time to cure identification (if cured), and $q_i$ the vector of characteristics associated with the latter.

Considering the time to cure identification, it follows that the observed likelihood is given by:

$$L_o = \prod_{i=1}^{n} p_Y(z_i;\beta)^{y_i\delta_i} f_T(t,x;\gamma)^{y_i\delta_i}\Big(1 \\
- p_Y(z_i;\beta)\Big)^{(1-y_i)\delta_i} h_c(t_i,q_i;\theta)^{(1-y_i)\delta_i} S_c(t_i,q_i;\theta)^{(1-y_i)(1-\delta_i)} \Big(p_y(z_i;\beta)S_c(t_i,q_i;\theta) \\
+ \big(1 - p_Y(z_i;\beta)\big)S_T(t,x;\gamma)\Big)^{(1-\delta_i)}$$

whereas the complete likelihood function is given by:

$$L_c(\beta,\gamma,\theta) = \prod_{i=1}^{n} p_Y(z_i;\beta)^{y_i} h_T(t_i,x_i;\gamma)^{y_i\delta_i} S_T(t_i,x_i;\gamma)^{y_i}\Big(1 \\
- p_Y(z_i;\beta)\Big)^{(1-y_i)} h_c(t_i,q_i;\theta)^{(1-y_i)\delta_i} S_c(t_i,q_i;\theta)^{(1-y_i)(1-\delta_i)}$$

where $h_c$ and $S_c$ the hazard and survival functions for the cure identification.

We need to point out here that $S_c$ is not the survival function for a cured individual; for them, the survival function is always equal to 1, since they will never experience the event under consideration. Instead, it represents the time of the cure identification, given that this subject belongs to the cure population. For situations where this is never observed or there is a specific timepoint after which all the individuals are cured, $S_c(t, \boldsymbol{q}; \boldsymbol{\theta}) = 1 \ \forall \ t, \boldsymbol{x}$. However, when the time to cure is stochastic (as in hospital stay), then $S_c$ can be any survival function. Lastly, this function can be assumed to be proper (i.e. $\lim_{t \to \infty} S_c(t, \boldsymbol{q}; \boldsymbol{\theta}) = 0$) since if the follow-up time could be infinite, then the cure status would always be identifiable.

The M-step of the algorithm seeks to maximize the function $El_c$, with

$$El_c(\boldsymbol{\beta}, \boldsymbol{\gamma}, \boldsymbol{\theta}) = \sum_{i=1}^{n} \left[ w_i \log(p_Y(\boldsymbol{z}_i; \boldsymbol{\beta})) + (1 - w_i) \log(1 - p_Y(\boldsymbol{z}_i; \boldsymbol{\beta})) \right]$$
$$+ \sum_{i=1}^{n} \left[ w_i \delta_i \log(h_T(t_i, \boldsymbol{x}_i; \boldsymbol{\gamma})) + w_i \log(S_T(t_i, \boldsymbol{x}_i; \boldsymbol{\gamma})) \right]$$
$$+ \sum_{i=1}^{n} \left[ (1 - w_i) \delta_i \log(h_c(t_i, \boldsymbol{q}_i; \boldsymbol{\theta})) + (1 - w_i) \log(S_c(t_i, \boldsymbol{q}_i; \boldsymbol{\theta})) \right]$$

To make it feasible for the model to distinguish between the identified immunes and the censored individuals, one needs to observe that, in that case, the constraint $\delta_i \leq y_i$ (4) does not hold, since when an individual is known to be cured, then $y_i = 0$ and $\delta_i = 1$. Therefore, the weight $w_i = 0$ should be assigned, leading to $w_i \delta_i = 0 \neq \delta_i = 1$. Moreover, for a censored subject, using Bayes Theorem, the probability to be uncured is given by:

$$P(Y_i = 1 | \delta_i = 0; t_i, \boldsymbol{x}_i, \boldsymbol{z}_i, \boldsymbol{q}_i) = \frac{P(\delta_i = 0 | Y_i = 1; t_i, \boldsymbol{x}_i, \boldsymbol{z}_i, \boldsymbol{q}_i) P(Y_i = 1; \boldsymbol{x}_i, \boldsymbol{z}_i, \boldsymbol{q}_i)}{P(\delta_i = 0; t_i, \boldsymbol{x}_i, \boldsymbol{z}_i, \boldsymbol{q}_i)}$$
$$= \frac{S_T(t_i, \boldsymbol{x}_i; \boldsymbol{\gamma}) p_Y(\boldsymbol{z}_i; \boldsymbol{\beta})}{(1 - p_Y(\boldsymbol{z}_i; \boldsymbol{\beta})) S_c(t_i, \boldsymbol{q}_i; \boldsymbol{\theta}) + p_Y(\boldsymbol{z}_i; \boldsymbol{\beta}) S_T(t_i, \boldsymbol{x}_i; \boldsymbol{\gamma})}$$

Hence, it turns out that the appropriate way is to assign the following weights:

$$w_i = \delta_i y_i + (1 - \delta_i) \frac{p_Y(\boldsymbol{z}_i; \boldsymbol{\beta}) S_T(t_i, \boldsymbol{x}_i; \boldsymbol{\gamma})}{(1 - p_Y(\boldsymbol{z}_i; \boldsymbol{\beta})) S_c(t_i, \boldsymbol{q}_i; \boldsymbol{\theta}) + p_Y(\boldsymbol{z}_i; \boldsymbol{\beta}) S_T(t_i, \boldsymbol{x}_i; \boldsymbol{\gamma})} \quad (5)$$

with

$$y_i = \begin{cases} 1, & \text{for known susceptible individuals} \\ 0, & \text{for known cured individuals} \\ -, & \text{for censored individuals} \end{cases}$$

Clearly, this model can cover all the three mechanisms that can result in cure status identification. Concerning the first mechanism (specific deterministic cutoff point, $c$), all the censored cases are associated with observed times lower than $c$, so $S_c(t_i, \boldsymbol{q}_i; \boldsymbol{\theta}) = 1$. Regarding the second mechanism, $S_c(t_i, \boldsymbol{q}_i; \boldsymbol{\theta})$ could be every proper survival function. As for the third mechanism, $S_c$ is actually a survival function of a discrete random variable, independent of the follow-up time, but dependent on $\boldsymbol{q}_i$. In other words, $S_c$ represents the survival function of a Bernoulli distribution with a success probability dependent on $\boldsymbol{q}_i$.

### 3.2. Properties of the new model

Before proceeding to the M-step of the EM algorithm, we present some important remarks that demonstrate the value of a model that manages to distinguish the identified immunes from the censored individuals. First, notice that for a sample with no known cured individuals, then $y_i = 1, \forall i: \delta_i = 1$, and the proposed model turns to the classic MC models proposed in the literature. Therefore, the new model constitutes a generalization of those models. However, when known cured individuals appear in the sample, since $S_c \leq 1$, the weights assigned to the censored cases (neither known cured nor susceptible) by the new model are now higher. Therefore, the proposed model gives, in general, a larger probability of a censored case to be susceptible, with the amount of increase depending on the distribution of the time to cure given cured and the censoring time.

Next, we examine each mechanism distinctively. Regarding the mechanism 1 (specific cutoff point after which all inexperienced individuals are cured), $S_c = 1 \forall t$. As a result, the weights assigned to the censored cases are identical to the ones of the hitherto MC models in the literature. This was also shown in Safari et al., (2022, 2021) for their respective nonparametric crude cure probability estimators. Concerning the mechanism 2 (stochastic time to cure identification), to better understand the "cost" of ignoring the cure status, assume that the population consists of cured individuals, but all are treated as censored (i.e. $y_i \delta_i = 0 = \delta_i$). In this case, although we should assign $w_i = 0$, we will instead use:

$$w_i = \frac{p_Y(\mathbf{z}_i; \boldsymbol{\beta}) S_T(t_i, \mathbf{x}_i; \boldsymbol{\gamma})}{1 - p_Y(\mathbf{z}_i; \boldsymbol{\beta}) + p_Y(\mathbf{z}_i; \boldsymbol{\beta}) S_T(t_i, \mathbf{x}_i; \boldsymbol{\gamma})} = \frac{S_T(t_i, \mathbf{x}_i; \boldsymbol{\gamma})}{\frac{1}{p_Y(\mathbf{z}_i; \boldsymbol{\beta})} - 1 + S_T(t_i, \mathbf{x}_i; \boldsymbol{\gamma})}$$

$$= \frac{p_Y(\mathbf{z}_i; \boldsymbol{\beta})}{\frac{1 - p_Y(\mathbf{z}_i; \boldsymbol{\beta})}{S_T(t_i, \mathbf{x}_i; \boldsymbol{\gamma})} + p_Y(\mathbf{z}_i; \boldsymbol{\beta})}$$

So, if the characteristics $\mathbf{z}_i$ are such that $p_Y(\mathbf{z}_i; \boldsymbol{\beta})$ becomes high, then $w_i$ diverges from 0 and tends to 1. The same also holds, if either the characteristics $\mathbf{x}_i$ or the $t_i$ are such that $S_T(t_i, \mathbf{x}_i; \boldsymbol{\gamma})$ becomes high (for the $t_i$ this means a low censoring/immune time). Hence, even though $w_i = 0$ should be assigned for the cases with small, realized curation time, the traditional MC models would use a $w_i$ closer to 1.

Lastly, under the mechanism 3 (i.e. $S_c(t)$ is replaced with a constant, say, $(1 - p_{obs})$, independent of $t$), then the Xue et al. (2022) model prevails. This concerns situations where a proportion of the individuals underlie a diagnostic test, independently of their follow-up time and should only be used for those. Otherwise, since $S_c(t)$ is a decreasing function of $t$ (as a survival function) while $p_{obs}$ is constant, the model of Xue et al. (2022) would assign lower (higher) weights in subjects with lower (higher) censoring times. On the other hand, Xue et al. (2022) model could be used as alternative when some cured individuals are known, but there not enough information that allows as precise estimation of the time until that happens. This is further discussed in Section 6.

## 4. Optimization

In this section, we will study the optimization process. To this end, and since the EM algorithm is solved iteratively, a method for $w_i$ initialization, a method for maximizing $El'_c$ for the given $w_i$ and a method for updating $w_i$ must be provided. We will first briefly present the maximization algorithm as proposed in the literature for the latency part (section 4.1), and then we will only describe the $w_i$ initialization and propose a method for updating $w_i$ (section 4.2).

### 4.1 Optimal formulas for the given $w_i$

Assume that at the $j$-th iteration of the EM algorithm, the weight $w_i^{(j)}$ has been assigned to the individual $i$. Then, if the population consists of both susceptible and immune individuals, in order to maximize $El'_c$, one can maximize the functions $El'_{c_1}$, $El'_{c_2}$ and $El'_{c_3}$, where:

$$El'_{c_1}(j) = \sum_{i=1}^{n} \left[ w_i^{(j)} \log\left(\frac{p_Y(\mathbf{z}_i; \boldsymbol{\beta})}{1 - p_Y(\mathbf{z}_i; \boldsymbol{\beta})}\right) + \log(1 - p_Y(\mathbf{z}_i; \boldsymbol{\beta})) \right]$$

$$El'_{c_2}(j) = \sum_{i=1}^{n} \left[ \delta_i w_i^{(j)} \log h_T(t_i, \mathbf{x}_i; \boldsymbol{\gamma}) + w_i^{(j)} \log(S_T(t_i, \mathbf{x}_i; \boldsymbol{\gamma})) \right]$$

$$El'_{c_3}(j) = \sum_{i=1}^{n} \left[ \delta_i \left(1 - w_i^{(j)}\right) \log h_C(t_i, \mathbf{x}_i; \boldsymbol{\gamma}) + \left(1 - w_i^{(j)}\right) \log(S_C(t_i, \mathbf{x}_i; \boldsymbol{\gamma})) \right]$$

Regarding the incidence part, the function $El'_{c_1}$ is similar to the ones used in the generalized linear models (see McCullagh and Nelder, 1989), therefore the optimization process has already been studied extensively in the literature. The same also holds for $El'_{c_2}(j)$, for a wide plethora of properties concerning the $S_T$ (e.g. semi-parametric with proportional hazards, Weibull, Exponential, etc.). For the special case of the semi-parametric with proportional hazards $S_T$, the function $El'_{c_2}$ is identical to the traditional Cox PH model, in which the term $\log(w_i)$ is added as an offset variable.

As for the identification part, since there are three different mechanisms that would result in different survival functions, they will be studied distinctively.

**Mechanism 1:** *Deterministic cutoff point after which all subjects are cured.*

In this case, $S_C(t_i, q_i; \theta) = 1$ for all censored cases. Therefore, the M-step of algorithm only needs to maximize the functions $El'_{c_1}$ and $El'_{c_2}$, and then update the weights using

$$w_i = \delta_i y_i + (1 - \delta_i) \frac{p_Y(z_i; \beta) S_T(t_i, x_i; \gamma)}{(1 - p_Y(z_i; \beta)) + p_Y(z_i; \beta) S_T(t_i, x_i; \gamma)}$$

**Mechanism 2:** *Stochastic time to cure identification.*

In this case, the function $S_C(t_i, q_i; \theta)$ can be any proper survival function. Therefore, the optimization processes for the $El'_{c_2}(j)$ apply for the $El'_{c_3}(j)$ maximization as well. For the special case of the semi-parametric with proportional hazards $S_c$, the function $El'_{c_3}$ is identical to the traditional Cox PH model, in which the term $\log(1 - w_i)$ is added as an offset variable.

The weights update is done using

$$w_i = \delta_i y_i + (1 - \delta_i) \frac{p_Y(z_i; \beta) S_T(t_i, x_i; \gamma)}{(1 - p_Y(z_i; \beta)) S_C(t_i, q_i; \theta) + p_Y(z_i; \beta) S_T(t_i, x_i; \gamma)}$$

**Mechanism 3:** *Diagnostic Procedure – Cure identification independent of the follow-up time.*

In this case, the function $S_C(t_i, q_i; \theta)$ can be derived from a Bernoulli distribution, therefore if $p_{obs}$ is the probability a cured subject to be observed, then in order to maximize $El'_{c_3}(j)$ one has to maximize

$$El'_{c_3}(j) = \sum_{\substack{i=1 \\ w_i^{(j)} \neq 1}}^{n} \left[ \left(1 - w_i^{(j)}\right) \log\left(\frac{p_{obs}(q_i; \theta)}{1 - p_{obs}(q_i; \theta)}\right) + \log(1 - p_{obs}(q_i; \theta)) \right]$$

which belongs to the traditional generalized linear models as the function $El'_{c_1}(j)$.

The weights update is done by using

$$w_i = \delta_i y_i + (1 - \delta_i) \frac{p_Y(z_i; \beta) S_T(t_i, x_i; \gamma)}{(1 - p_Y(z_i; \beta))(1 - p_{obs}(q_i; \theta)) + p_Y(z_i; \beta) S_T(t_i, x_i; \gamma)}$$

**4.2 Method for $w_i$ initialization and update**

It is obvious that to implement the EM algorithm, one needs to apply a method that provides initial values for the $w_i$, as well as the way to update them. For the former, an easily applicable and commonly suggested method is to firstly ignore the censorship. Therefore, at EM initialization, $w_i = \delta_i y_i$, which means $w_i = 1$ for all the uncensored susceptible individuals, and $w_i = 0$, otherwise.

Concerning the update of the $w_i$, it can be done directly after the completion of each step optimization process. Therefore, we need to estimate the $S_T$ and $S_c$, and consequently $h_T(t; 0)$ and $h_C(t; 0)$. To this end, we will use a Breslow-type method proposed by Sy and Taylor (2000) based on the profile likelihood. It is a modification of the Aalen-Nelson estimator and, for $h_T(t; 0)$ is given by:

$$\widehat{H}_{0T}(t) = \sum_{i:t_{(i)}\leq t} \left( \frac{d_i}{\sum_{m\in R_i} w_m^{(j)} e^{\gamma' x_m}} \right)$$

where $H_{0T}(t) = \int_0^t h_T(u; 0) du$, $t_{(i)}$ is the $i$-th order observed time, $d_i$ the number of events at time $t_{(i)}$, and $R_i$ the set containing the cases at risk at time $t_{(i)}^-$.

The function $\widehat{H}_{0T}(t)$ is then substituted to $El'_{c_2}$, and the maximization with respect to $\gamma$ is performed.

The question that arises here is which cases constitute the risk set $R_i$. Clearly, since the hazard function concerns only the susceptible population, the individuals that are known to be cured should be excluded from the risk set. Notice, however, that the formula $\sum_{m\in R_i} w_m^{(j)} e^{\gamma' x_m}$ assigns to each subject its weight, $w_m^{(j)}$. For the uncensored immunes, $w_m^{(j)} = 0$, so their inclusion in the risk set makes no difference.

The same approach is followed for the time to cure identification, now using $1 - w_m^{(j)}$ instead of $w_m^{(j)}$.

## 5. The penalized MC model with known cured individuals for variable selection

In this section, we apply penalized regression methods such as the least absolute shrinkage and selection operator (LASSO) in order to allow variable selection during the maximization of the loglikelihood function. The proposed method is a direct extension of the one studied by Liu *et al.* (2012), for the respective MC model with no observed cured individuals.

$$El_c(\boldsymbol{\beta},\boldsymbol{\gamma},\boldsymbol{\theta}) = \sum_{i=1}^n \left[w_i \log(p_Y(z_i;\boldsymbol{\beta})) + (1-w_i)\log(1-p_Y(z_i;\boldsymbol{\beta}))\right]$$
$$+ \sum_{i=1}^n \left[w_i \delta_i \log(h_T(t_i,x_i;\boldsymbol{\gamma})) + w_i \log(S_T(t_i,x_i;\boldsymbol{\gamma}))\right]$$
$$+ \sum_{i=1}^n \left[(1-w_i)\delta_i \log(h_c(t_i,q_i;\boldsymbol{\theta})) + (1-w_i)\log(S_c(t_i,q_i;\boldsymbol{\theta}))\right] - n\sum_{i=1}^{n_\beta} p_{\lambda_{1i}}(|\beta_i|)$$
$$- n\sum_{i=1}^{n_\gamma} p_{\lambda_{2i}}(|\gamma_i|) - n\sum_{i=1}^{n_\theta} p_{\lambda_{3i}}(|\theta_i|)$$

with $p_\lambda(|\cdot|)$ a penalty function, $\boldsymbol{\lambda} = \left(\lambda_{11}, \ldots, \lambda_{1n_\beta}, \lambda_{21}, \ldots, \lambda_{2n_\gamma}, \lambda_{31}, \ldots, \lambda_{3n_\theta}\right)'$ a tuning parameter vector and $n_\beta, n_\gamma$ and $n_\theta$ the length of the vectors $\boldsymbol{\beta}, \boldsymbol{\gamma}$ and $\boldsymbol{\theta}$, respectively.

Therefore, the objective is the maximization of the function $El'_{c_1}, El'_{c_2}$ and $El'_{c_3}$ where:

$$El'_{c_1}(\boldsymbol{\beta}) = \sum_{i=1}^n \left[w_i \log(p_Y(z_i;\boldsymbol{\beta})) + (1-w_i)\log(1-p_Y(z_i;\boldsymbol{\beta}))\right] - n\sum_{i=1}^{n_\beta} p_{\lambda_{1i}}(|\beta_i|)$$

$$El'_{c_2}(\boldsymbol{\gamma}) = \sum_{i=1}^n \left[w_i \delta_i \log(h_T(t_i,x_i;\boldsymbol{\gamma})) + w_i \log(S_T(t_i,x_i;\boldsymbol{\gamma}))\right] - n\sum_{i=1}^{n_\gamma} p_{\lambda_{2i}}(|\gamma_i|)$$

$$El'_{c_3}(\boldsymbol{\theta}) = \sum_{i=1}^n \left[(1-w_i)\delta_i \log(h_c(t_i,q_i;\boldsymbol{\theta})) + (1-w_i)\log(S_c(t_i,q_i;\boldsymbol{\theta}))\right] - n\sum_{i=1}^{n_\theta} p_{\lambda_{3i}}(|\theta_i|)$$

which can be achieved by using already existing penalized procedures (e.g. for the logistic and the Cox PH).

## 6. Simulation study

**6.1 Data generation**

In this section, we present the results of the simulation study conducted. Recall that in Sections 3.2 and 4.1, we showed that the MC model under the mechanism 1 (deterministic cutoff) is equivalent to the traditional MC models that ignore the known cured status. Therefore, the simulations only concerned the other two mechanisms for cure identification: (1) stochastic time to cure identification, (2) diagnostic test. The objective was to evaluate the proposed model, both in absolute terms and relative to previous existing models in the literature. Specifically, the following models were compared:

1) Model that considers all the information regarding the cure status, including the time until that happened, if available.
2) Model that only considers the cure status information, ignoring the time until that happened (if available) (Xue *et al.*, 2022).
3) Model that assigns infinite time-to-event times to the known cured subjects and treats them as censored.
4) Model that completely ignores the cure status (traditional models).

and the metrics used for the comparisons were:

1) The estimate and the 95% bootstrap confidence intervals (95% CI) based on 100 bootstrapping samples.
2) The mean square error (MSE).
3) The coverage probability (CP) based on 100 simulated datasets.

For the data generation, we used the following scenarios:

1) We constructed six independent variables: three binary and three continuous. The binary followed a Bernoulli distribution, whereas the continuous the Standard Normal.
2) We used four independent variables (two binary and two continuous) for each part (incidence and latency), such that exactly one binary and one continuous to be common between the two parts. The incidence part followed the logit model, whereas the latency was constructed through the proportional hazards assumption.
3) The baseline survival functions for the time to event and time to cure identification were both derived from the Weibull distribution.
4) We derived the censoring times using the Exponential distribution, with parameters such that the prespecified censoring proportions to be achieved.

To examine a great variety of the comparisons, two different scenarios regarding the stochastic time to cure identification were constructed stochastically lower than the time to event occurrence or stochastically higher. The respective distributions for the two scenarios are depicted in the Figure 1 and Figure 2.

The code for the simulation implementation, along with other functions for adjusting semi-parametric or fully-parametric models, or even models without independent variables for the time to cure identification is written in R and is available at Github (https://github.com/gkarakatsoulis/MCknownCured).

A sample of the results obtained is presented below.

**6.2 Simulation Results**

We start by presenting the results for mechanism 2 (stochastic time to cure identification). As previously mentioned, we have considered two scenarios regarding the distributions of the time to cure identification (for the cured) and time to event occurrence (for the susceptible): one where the former is stochastically lower than the latter (Figure 1) and one where it is stochastically higher (Figure 2). For both scenarios, we investigated both low and high cure rates (~10% and ~30%, respectively).

Table 1 and Table 2 display the results for low (~10%) and high (~30%) cure rates, respectively, when the time to cure identification is stochastically lower than the time to event occurrence. From both tables we observe a bias reduction as the sample size increases, along with a decreased MSE and increased precision (decreased confidence interval length). On the contrary, the CP does not display this behavior, a result that may be due to the large MSE when the sample size is small, and even the number of bootstrapping samples.

Regarding the comparison between the new model and the one that ignores the cure information, although the biases may be comparable, a result also shown in Safari *et al.* (2021), the MSEs in the former are considerably lower. Notice that regarding the incidence part, the traditional model displays huge MSEs, especially when the sample size is small, which almost diminish with the new model.

Tables Table 3 and Table 4 display the results for low (~10%) and high (~30%) cure rates, respectively, when the time to cure identification is stochastically higher than the time to event occurrence. As expected, the traditional cure models perform better compared to the cases with lower time to cure identification. However, we still observe situations with huge MSEs that almost diminish with the new model. Notice, here, that Table Table 3 does not include an estimation with sample size equal to 250. This is due to the very low number of known cured individuals appearing to the dataset (0 in some cases), which forbids us to estimate the time until cure identification.

The incapability of yielding estimates when the number of known cured subjects is very low motivated us to explore two other strategies for including the cure information, without the need for estimating the time to cure identification, when it is not possible. Both strategies rely on the traditional mixture cure models, with a slight modification either in the weight assignment or in the observed times for the known cured individuals. The first strategy (crude cure probability) is the model of Xue *et al.* (2022). The second strategy assigns very large censoring times to the known cured individuals (higher than the max observed time), treats them as censored, and then adjusts a traditional mixture cure model. The results are displayed in Figure 3. The true parameter values are given in Table 4, sample size 250. As it can be seen, the "Crude cure probability" and the "Time-to-event infinite" strategies perform considerably better than the traditional mixture cure models that ignore this information, and are comparable to the model that considers the time-to-cure information. Therefore, those strategies could be use as alternatives when the number of known cured individuals does now allow a precise estimation for the time until curation is achieved.

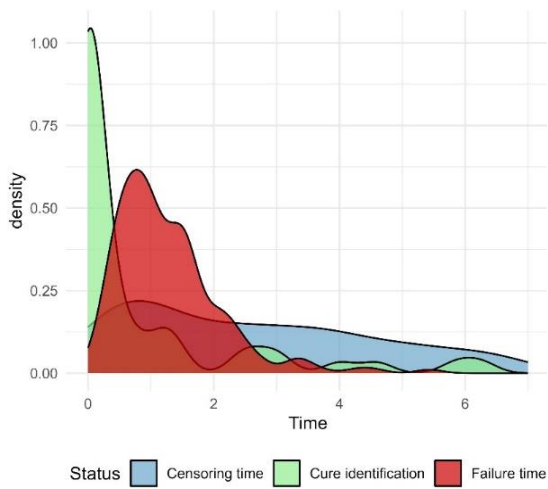

*Figure 1. Distribution of true times to event for the population distribution with times until cure identification being stochastically lower than the times until the failure.*

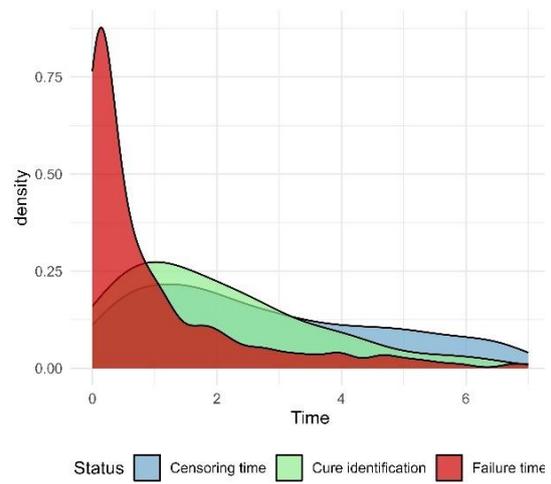

*Figure 2. Distribution of true times to event for the population distribution with times until cure identification being stochastically higher than the times until the failure.*

| | | | Low cure rate (~10%) – High Known Cured Rate (~50%) | | | | | |
|---|---|---|---|---|---|---|---|---|
| | | | Cure information | | | | | |
| | | | Consider (New model) | | | Ignore (Traditional models) | | |
| N | Coefficient | True | Est. (95% CI) | MSE | CP | Est. (95% CI) | MSE | CP |
| 250 | $\beta_0$ | 2 | 2.42 (2.28, 2.56) | 0.67 | 78 | 32.67 (-8.25, 73.58) | 44076 | 56 |
| | $\beta_1$ | 1 | 0.98 (0.9, 1.06) | 0.16 | 95 | 9.89 (-11.51, 31.28) | 11877 | 88 |
| | $\beta_2$ | 2 | 3.34 (2.49, 4.19) | 20.24 | 86 | 17.04 (-5.28, 39.36) | 13064 | 85 |
| | $\beta_3$ | 1 | 1.1 (1.02, 1.18) | 0.17 | 89 | 30.81 (-23.72, 85.33) | 77515 | 93 |
| | $\beta_4$ | 0.5 | 0.38 (0.24, 0.52) | 0.51 | 95 | 20.41 (-16.38, 57.2) | 35275 | 90 |
| | $\gamma_1$ | 0.9 | 0.9 (0.87, 0.92) | 0.01 | 96 | 0.93 (0.91, 0.95) | 0.01 | 95 |
| | $\gamma_2$ | 1 | 0.93 (0.88, 0.98) | 0.06 | 97 | 1.01 (0.98, 1.05) | 0.03 | 97 |
| | $\gamma_3$ | 4 | 3.77 (3.68, 3.87) | 0.3 | 97 | 4.06 (4.01, 4.11) | 0.07 | 92 |
| | $\gamma_4$ | 2 | 1.9 (1.82, 1.97) | 0.15 | 95 | 2.06 (2.01, 2.11) | 0.07 | 92 |
| 500 | $\beta_0$ | 2 | 2.34 (2.22, 2.46) | 0.48 | 73 | 4.01 (3.28, 4.75) | 18.07 | 55 |
| | $\beta_1$ | 1 | 1.03 (0.97, 1.08) | 0.09 | 93 | 1.14 (1.05, 1.23) | 0.24 | 90 |
| | $\beta_2$ | 2 | 2.08 (1.95, 2.21) | 0.44 | 89 | 5.07 (3.78, 6.37) | 52.75 | 91 |
| | $\beta_3$ | 1 | 1.01 (0.97, 1.06) | 0.06 | 95 | 1.1 (0.97, 1.23) | 0.44 | 91 |
| | $\beta_4$ | 0.5 | 0.49 (0.37, 0.61) | 0.35 | 88 | -0.14 (-0.87, 0.6) | 14.45 | 95 |
| | $\gamma_1$ | 0.9 | 0.87 (0.85, 0.89) | 0.01 | 94 | 0.91 (0.9, 0.93) | 0.01 | 88 |
| | $\gamma_2$ | 1 | 0.92 (0.88, 0.96) | 0.05 | 90 | 1.02 (0.99, 1.04) | 0.02 | 95 |
| | $\gamma_3$ | 4 | 3.69 (3.59, 3.8) | 0.38 | 91 | 4.05 (4.01, 4.08) | 0.04 | 90 |
| | $\gamma_4$ | 2 | 1.83 (1.77, 1.9) | 0.14 | 92 | 2.01 (1.98, 2.04) | 0.03 | 91 |

Table 1. Comparison of the proposed model with the one that ignores the curation status and the time to cure identification, when the latter is stochastic (mechanism 2). Times until cure identification stochastically lower than the times until the failure.

| | | | High cure rate (~30%) – High Known Cured Rate (~50%) | | | | | |
|---|---|---|---|---|---|---|---|---|
| | | | Cure information | | | | | |
| | | | Consider (New model) | | | Ignore (Traditional models) | | |
| N | Coefficient | True | Est. (95% CI) | MSE | CP | Est. (95% CI) | MSE | CP |
| 250 | $\beta_0$ | 2 | 2.4 (2.2, 2.59) | 1.14 | 89 | 10.62 (2.29, 18.95) | 1862 | 69 |
| | $\beta_1$ | 4 | 4.24 (4, 4.48) | 1.56 | 93 | 14.3 (2.93, 25.67) | 3439 | 80 |
| | $\beta_2$ | 2 | 2.12 (1.96, 2.28) | 0.68 | 96 | 10.74 (-1.49, 22.98) | 3934 | 88 |
| | $\beta_3$ | 4 | 4.21 (3.97, 4.45) | 1.51 | 94 | 16.13 (2.22, 30.03) | 5130 | 77 |
| | $\beta_4$ | 0.5 | 0.41 (0.24, 0.58) | 0.72 | 93 | 0.28 (-2.21, 2.77) | 160 | 94 |
| | $\gamma_1$ | 0.9 | 0.92 (0.9, 0.95) | 0.02 | 88 | 0.94 (0.92, 0.96) | 0.02 | 90 |
| | $\gamma_2$ | 1 | 0.97 (0.93, 1.02) | 0.05 | 96 | 1.01 (0.97, 1.05) | 0.04 | 96 |
| | $\gamma_3$ | 4 | 3.96 (3.86, 4.05) | 0.22 | 92 | 4.08 (4.01, 4.14) | 0.12 | 88 |
| | $\gamma_4$ | 2 | 1.99 (1.92, 2.05) | 0.12 | 94 | 2.06 (2, 2.11) | 0.08 | 88 |
| 500 | $\beta_0$ | 2 | 2.21 (2.11, 2.31) | 0.3 | 0.91 | 3.28 (3.09, 3.48) | 2.64 | 0.71 |
| | $\beta_1$ | 4 | 4.18 (4.03, 4.32) | 0.6 | 0.94 | 4.52 (4.3, 4.73) | 1.44 | 0.89 |
| | $\beta_2$ | 2 | 2.21 (2.09, 2.33) | 0.38 | 0.93 | 2.34 (2.16, 2.51) | 0.91 | 0.92 |
| | $\beta_3$ | 4 | 4.17 (4.02, 4.32) | 0.57 | 0.92 | 4.41 (4.22, 4.6) | 1.11 | 0.89 |
| | $\beta_4$ | 0.5 | 0.4 (0.31, 0.5) | 0.23 | 0.96 | 0.37 (0.2, 0.54) | 0.78 | 0.92 |
| | $\gamma_1$ | 0.9 | 0.9 (0.89, 0.92) | 0.01 | 0.93 | 0.91 (0.9, 0.93) | 0.01 | 0.94 |
| | $\gamma_2$ | 1 | 1 (0.97, 1.04) | 0.03 | 0.94 | 1.02 (0.99, 1.05) | 0.02 | 0.9 |
| | $\gamma_3$ | 4 | 3.98 (3.92, 4.05) | 0.12 | 0.94 | 4.05 (4, 4.09) | 0.05 | 0.91 |
| | $\gamma_4$ | 2 | 1.98 (1.94, 2.02) | 0.05 | 0.92 | 2.01 (1.97, 2.04) | 0.03 | 0.91 |

Table 2. Comparison of the proposed model with the one that ignores the curation status and the time to cure identification, when the latter is stochastic (mechanism 2). Times until cure identification stochastically lower than the times until the failure.

| | | | Low cure rate (~10%) – High Known Cured Rate (~50%) | | | | | |
|---|---|---|---|---|---|---|---|---|
| | | | Cure information | | | | | |
| | | | Consider (New model) | | | Ignore (Traditional models) | | |
| N | Coefficient | True | Est. (95% CI) | MSE | CP | Est. (95% CI) | MSE | CP |
| 500 | $\beta_0$ | 2 | (1.92, 2.11) | 0.21 | 96 | (2.54, 3.31) | 4.7 | 96 |
| | $\beta_1$ | 1 | (1.01, 1.12) | 0.09 | 88 | (0.92, 1.1) | 0.2 | 91 |
| | $\beta_2$ | 2 | (1.98, 2.2) | 0.33 | 92 | (2.19, 4.02) | 22.66 | 85 |
| | $\beta_3$ | 1 | (1.04, 1.14) | 0.07 | 93 | (0.87, 1.12) | 0.39 | 79 |
| | $\beta_4$ | 0.5 | (0.46, 0.68) | 0.32 | 83 | (-0.06, 0.69) | 3.7 | 97 |
| | $\gamma_1$ | 0.9 | (0.89, 0.91) | <0.001 | 95 | (0.89, 0.91) | <0.001 | 93 |
| | $\gamma_2$ | 1 | (0.97, 1.03) | 0.03 | 89 | (1, 1.05) | 0.02 | 92 |
| | $\gamma_3$ | 4 | (3.92, 4.05) | 0.11 | 87 | (4, 4.07) | 0.04 | 89 |
| | $\gamma_4$ | 2 | (1.94, 2.03) | 0.05 | 92 | (1.98, 2.04) | 0.03 | 90 |

Table 3. Comparison of the proposed model with the one that ignores the curation status and the time to cure identification, when the latter is stochastic (mechanism 2). Times until cure identification stochastically higher than the times until the failure.

| | | | High cure rate (~30%) – High Known Cured Rate (~50%) | | | | | |
|---|---|---|---|---|---|---|---|---|
| | | | Cure information | | | | | |
| | | | Consider (New model) | | | Ignore (Traditional models) | | |
| N | Coefficient | True | Est. (95% CI) | MSE | CP | Est. (95% CI) | MSE | CP |
| 250 | $\beta_0$ | 2 | 2.18 (2, 2.37) | 0.92 | 91 | 23.47 (-16.71, 63.65) | 42066 | 97 |
| | $\beta_1$ | 4 | 4.22 (4, 4.44) | 1.3 | 96 | 34.38 (-23.97, 92.73) | 88656 | 96 |
| | $\beta_2$ | 2 | 2.12 (1.95, 2.29) | 0.79 | 97 | 5.94 (-1.09, 12.97) | 1289 | 97 |
| | $\beta_3$ | 4 | 4.36 (4.15, 4.58) | 1.32 | 96 | 27.12 (-17.21, 71.45) | 51174 | 88 |
| | $\beta_4$ | 0.5 | 0.54 (0.37, 0.71) | 0.72 | 97 | 9.82 (-8.43, 28.08) | 8676 | 93 |
| | $\gamma_1$ | 0.9 | 4.22 (4, 4.44) | 1.3 | 96 | 0.93 (0.91, 0.96) | 0.01 | 91 |
| | $\gamma_2$ | 1 | 2.12 (1.95, 2.29) | 0.79 | 97 | 1.02 (0.98, 1.05) | 0.04 | 95 |
| | $\gamma_3$ | 4 | 4.01 (3.95, 4.08) | 0.11 | 93 | 4.09 (4.03, 4.15) | 0.1 | 92 |
| | $\gamma_4$ | 2 | 2.01 (1.96, 2.06) | 0.07 | 94 | 2.05 (2, 2.1) | 0.07 | 87 |
| 500 | $\beta_0$ | 2 | 2.01 (1.91, 2.1) | 0.25 | 94 | 2.58 (2.39, 2.77) | 1.24 | 93 |
| | $\beta_1$ | 4 | 4.05 (3.92, 4.19) | 0.46 | 99 | 4.15 (3.96, 4.35) | 0.99 | 95 |
| | $\beta_2$ | 2 | 2.15 (2.04, 2.27) | 0.35 | 98 | 2.18 (2.01, 2.34) | 0.72 | 95 |
| | $\beta_3$ | 4 | 4.18 (4.05, 4.31) | 0.47 | 94 | 3.85 (3.7, 4) | 0.6 | 94 |
| | $\beta_4$ | 0.5 | 0.46 (0.37, 0.55) | 0.21 | 97 | 0.35 (0.2, 0.5) | 0.61 | 95 |
| | $\gamma_1$ | 0.9 | 0.91 (0.89, 0.92) | 0.01 | 92 | 0.91 (0.89, 0.92) | 0.01 | 92 |
| | $\gamma_2$ | 1 | 1 (0.97, 1.03) | 0.02 | 93 | 1.02 (0.99, 1.05) | 0.02 | 91 |
| | $\gamma_3$ | 4 | 3.99 (3.94, 4.03) | 0.05 | 95 | 4.06 (4.02, 4.1) | 0.05 | 90 |
| | $\gamma_4$ | 2 | 1.98 (1.94, 2.02) | 0.03 | 89 | 2.01 (1.98, 2.05) | 0.03 | 89 |

Table 4. Comparison of the proposed model with the one that ignores the curation status and the time to cure identification, when the latter is stochastic (mechanism 2). Times until cure identification stochastically higher than the times until the failure.

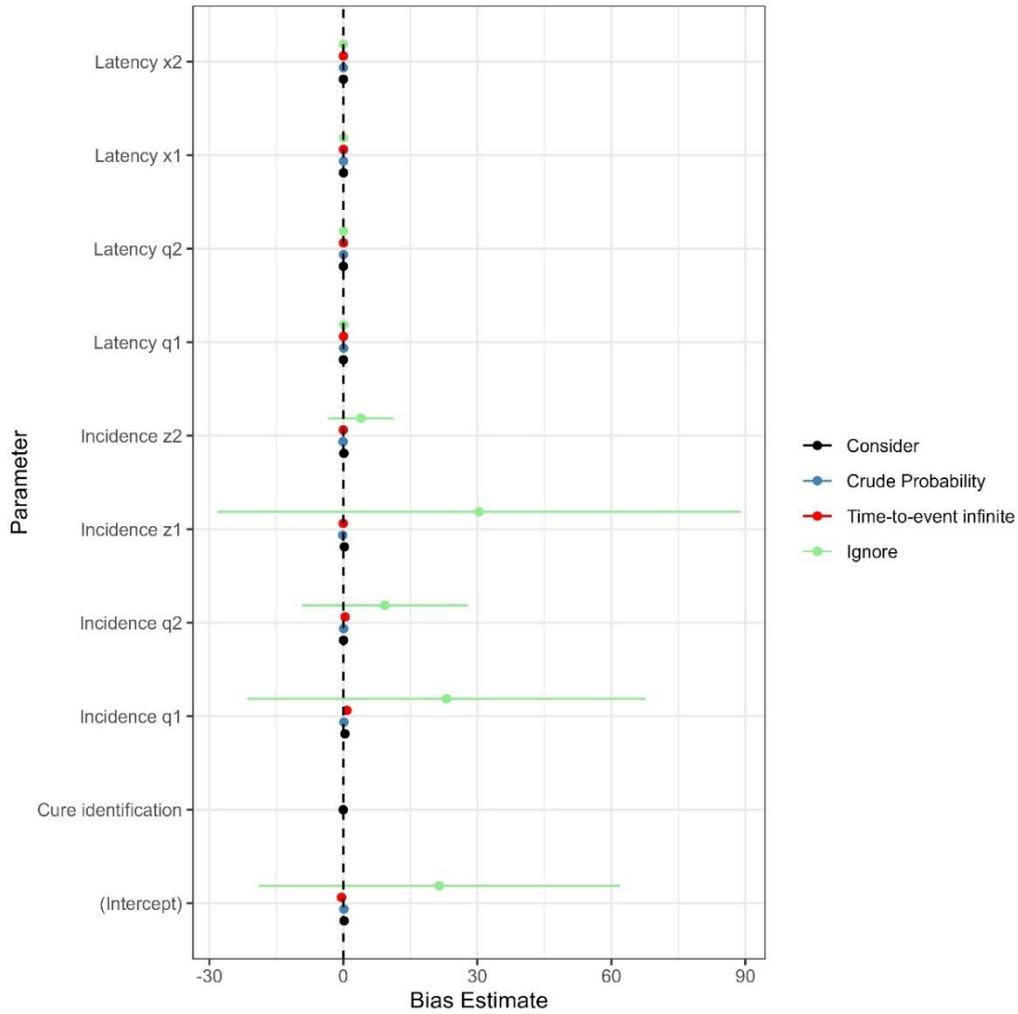

*Figure 3. Comparison of the different strategies to include the known cured information in the model.*

## 7. Discussion and Conclusion

In this study we propose a Mixture Cure model that incorporates information on known cured individuals, which is often ignored in the traditional models. There are several aspects that constitute a subject to be known cured: (1) deterministic cutoff after which all subjects are cured, (2) stochastic time to cure, (3) diagnostic procedures. Characteristic examples of those aspects constitute the x-year survival rate, the death during hospital stay (subjects that leave the hospital are cured) and the diagnostic tests conducted to the subjects, respectively.

To take advantage of such information, we build a new mathematical formulation and construct the complete loglikelihood, which we maximize through the EM algorithm to obtain the MLEs. This is also done under the LASSO penalty to allow for variable selection method. We then compare the new model with the traditional, both theoretically and numerically (through simulations), across diverse scenarios, including varying cure rates and sample sizes. Under the deterministic cutoff point after which all subjects are cured, the new model is equivalent to the traditional ones. Regarding the case with stochastic time to cure identification, the new model significantly improves the estimation, with simulated results demonstrating higher precision, higher coverage probability and lower mean squared errors. Notably, the last ones may be very high in the traditional models but almost diminished in the new, especially when the time to cure identification (for cured) is stochastically lower than the time to event occurrence (for the susceptible). This is expected due to the mechanism under which the traditional models consider a censored case as cured: they assign higher cure probabilities to cases with high censoring time. However, when the time to cure identification is stochastically lower than the time to event occurrence, the known cured subjects in general display low observed times.

Although the new model outperforms the traditional ones, it may not be applicable when the number of known cured subjects is low. Therefore, we investigated two other strategies for such cases. without the need for estimating the time to cure identification, when it is not possible. Both strategies rely on the traditional mixture cure models, with a slight modification either in the weight assignment or in the observed times for the known cured individuals. The first strategy (crude cure probability) is the model of Xue *et al.* (2022). The second strategy assigns very large censoring times to the known cured individuals (higher than the max observed time), treats them as censored, and then adjusts a traditional mixture cure model. The simulated results show that those strategies constitute robust alternatives when the number of known cured individuals is small, leading to more precise estimation compared to the traditional models.

Regarding future research, the known cured information could be used in more flexible families of Cure models (e.g. Promotion time model and Box-Cox model). Furthermore, since this model may share similarities with competing risk approaches, comparisons between those two could highlight the advantages of each approach.

**Conflict of Interest**

*The authors have declared no conflict of interest*